\documentclass[12pt]{article}
\usepackage{latexsym}
\usepackage{amsthm}
\usepackage{amsfonts,bm,amsbsy,bbm,amsmath}
\usepackage{epsfig}
\usepackage{graphicx,rotating,lscape}
\usepackage{verbatim}
\usepackage{natbib}
\usepackage{multirow,color}
\usepackage{geometry}
\usepackage{setspace}
\usepackage{subcaption}
\usepackage{float}
\usepackage{dcolumn}
\usepackage{siunitx}
\usepackage{algorithm}
\usepackage{longtable}
\usepackage{makecell}
\usepackage{enumerate}
\usepackage{booktabs}
\usepackage{authblk}
\usepackage{tikz}
  \usetikzlibrary{positioning,shapes.geometric}
\usepackage{hyperref}
\hypersetup{colorlinks=true,linkcolor=blue,urlcolor=blue,citecolor=blue}

\usepackage{cleveref}
\usepackage{algorithm}
\usepackage{algorithmic}
\usepackage{tcolorbox}
\tcbuselibrary{breakable}

\usepackage{amsthm,amssymb}
\newtheorem{theorem}{Theorem}

\newtheorem{assum}{Assumption}
\theoremstyle{definition}

\newtheorem{remark}{Remark}
\newtheorem{example}{Example}

\newtheoremstyle{exctd}
{\topsep} {\topsep}%
{\upshape}
{}
{\bfseries}
{.}
{5pt plus 1pt minus 1pt}
{\thmname{#1} \thmnumber{ #2}\thmnote{#3} (Cont'd)}
\theoremstyle{exctd}
\newtheorem*{exctd}{Example}

\crefname{theorem}{theorem}{theorems}
\crefname{cond}{condition}{conditions}
\crefname{assum}{assumption}{assumptions}
\crefname{lemma}{lemma}{lemmas}
\crefname{prop}{proposition}{propositions}
\crefname{coro}{corollary}{corollaries}

\Crefname{cond}{Condition}{Conditions}
\Crefname{theorem}{Theorem}{Theorems}
\Crefname{assum}{Assumption}{Assumptions}
\Crefname{lemma}{Lemma}{Lemmas}
\Crefname{prop}{Proposition}{Propositions}
\Crefname{coro}{Corollary}{Corollaries}

\crefrangelabelformat{assum}{#3#1#4--#5#2#6}
\crefrangelabelformat{section}{#3#1#4--#5#2#6}

\usepackage{multirow}

\def\beqr{\begin{eqnarray}}
\def\eeqr{\end{eqnarray}}
\def\beqrs{\begin{eqnarray*}}
\def\eeqrs{\end{eqnarray*}}

\def\mR{\mathbb{R}}

\def\calA{\mathcal{A}}
\def\calH{\mathcal{H}}
\def\calG{\mathcal{G}}
\def\calL{\mathcal{L}}
\def\calM{\mathcal{M}}
\def\calR{\mathcal{R}}

\def\calX{\mathcal{X}}

\def\calV{\mathcal{V}}
\def\fX{\mathfrak{X}}
\def\fx{\mathfrak{x}}
\def\fZ{\mathfrak{Z}}
\def\fz{\mathfrak{z}}

\def\X{\mathbf{X}}
\def\Z{\mathbf{Z}}
\def\R{\mathbf{R}}
\def\p{\mathbf{p}}
\def\x{\mathbf{x}}

\def\r{\mathbf{r}}
\def\1{\mathbf{1}}
\def\0{\mathbf{0}}
\def\O{\mathbf{O}}

\def\wh{\widehat}

\def\indep{\bot\!\!\!\bot}

\def\full{\text{\normalfont{full}}}
\def\loss{\text{\normalfont{loss}}}
\def\mis{\text{\normalfont{mis}}}

\newcommand{\MARKER}[1]{}

\newcommand{\trans}{^\top}
\newcommand{\Abs}[1]{\left\vert#1\right\vert}
\newcommand{\pprime}{^\prime}
 
\newcommand{\amin}{\operatornamewithlimits{arg\,min}}

\usepackage{xr}
\def\cf{\textnormal{cf}}

\makeatletter
\newcommand*{\ind}{%
\mathbin{%
\mathpalette{\@ind}{}%
}%
}
\newcommand*{\nind}{%
\mathbin{
\mathpalette{\@ind}{\not}
}%
}
\newcommand*{\@ind}[2]{%
\sbox0{$#1\perp\m@th$}
\sbox2{$#1=$}
\sbox4{$#1\vcenter{}$}
\rlap{\copy0}
\dimen@=\dimexpr\ht2-\ht4-.2pt\relax
\kern\dimen@
{#2}%
\kern\dimen@
\copy0 
}
\makeatother

\begin{document}

\title{Efficient Nonparametric Inference for Mediation Analysis with Nonignorable Missing Confounders}

\author[1]{Jiawei Shan}
\author[2]{Wei Li}
\author[3]{Chunrong Ai}
\affil[1]{Institute of Statistics and Big Data, Renmin University of China}
\affil[2]{Center for Applied Statistics and School of Statistics, Renmin University of China}
\affil[3]{School of Management and Economics, The Chinese University of Hong Kong, Shenzhen}

\date{\today}

\maketitle
\thispagestyle{empty}

\begin{abstract}
 Mediation analysis is widely used for exploring treatment mechanisms; however, it faces challenges when nonignorable missing confounders are present. Efficient inference of mediation effects and the efficiency loss due to nonignorable missingness have been rarely studied in the literature because of the difficulties arising from the ill-posed inverse problem. In this paper, we propose a general shadow variable framework for identifying mediation effects, allowing shadow variables to be selected from either observed covariates or externally collected auxiliary data. We then propose a Sieve-based Iterative Outward (SIO) approach for estimation. We establish large-sample theory, particularly asymptotic normality, for the proposed estimator despite the ill-posedness of the problem. We show that our estimator is locally efficient and attains the semiparametric efficiency bound under certain conditions. Building on the efficient influence function, we explicitly quantify the efficiency loss attributable to missingness and propose a debiased machine learning approach for estimation and inference. We examine the finite-sample performance of the proposed approach using extensive simulation studies and showcase its practical applicability through an empirical analysis of CFPS data.
\end{abstract}
{\bf Key Words:} Direct and indirect effects; Identification; Ill-posed inverse problem; Missing not at random; Semiparametric efficiency bound.

\newpage

\section{Introduction}

In scientific disciplines such as epidemiology, biostatistics, and the social sciences, researchers are increasingly interested in understanding the pathways by which a treatment or intervention influences an outcome. Causal mediation analysis is a powerful analytical approach that aims to elucidate the underlying mechanisms of treatment effects \citep{BaronKenny1986,RobinsGreenland1992,Pearl2001,ImaiKeeleTingley2010,ImaiKeeleYamamoto2010,TchetgenTchetgenShpitser2012}.
This approach typically relies on the sequential ignorability assumption proposed by \cite{ImaiKeeleYamamoto2010}, which assumes the treatment is ignorable given pre-treatment covariates, and  the mediator is ignorable given the treatment and pre-treatment covariates. The pre-treatment covariates, also known as confounders, play a vital role in addressing the endogeneity associated with both the treatment and the mediator. 
The presence of missing values in covariates is familiar due to accidents, high costs, or privacy concerns, and it introduces novel challenges for causal mediation analysis.

Our work is motivated by an empirical study investigating the mediating role of subjective well-being in the potential relationship between job satisfaction and depressive symptoms among Chinese adults aged 35-60 years. The analysis is based on data from the China Family Panel Studies (CFPS), which focuses on both economic and non-economic aspects of well-being, including topics such as economic activities, education outcomes, family dynamics, migration, and health in contemporary China. To address potential endogeneity, a set of individual characteristics is used as pre-treatment covariates, such as age, gender, family size, and income. However, mediation analysis is challenged by approximately 44\% missing values in personal income data.  The missingness is likely to be nonignorable \citep{Rubin1976}, in the sense that it depends on missing values even after conditioning on observed data. This is because individuals with high incomes are often less inclined to disclose their earnings.
Existing estimation methods for complete data, such as regression \citep{ValeriVanderweele2013}, weighting \citep{Huber2014}, imputation \citep{VansteelandtBekaertLange2012}, and multiply robust methods \citep{TchetgenTchetgenShpitser2012}, may yield biased estimates and incorrect inference. \cite{LiZhou2017} developed an approach for mediation analysis with nonignorable missing outcomes using instrumental-variable-type covariates; \cite{ZuoGhoshDingYang2024} studied the identification of causal mediation effects when the mediator, the outcome, or both are missing, by imposing a specific structure on the missingness mechanism. 
All these methods, however, are not applicable to address nonignorable missing covariates encountered in our empirical study.

\subsection{Prior work}\label{subsec:prior}
Missing covariates bring two primary challenges to causal mediation analysis.
The first challenge is identification, because the full data distribution and causal effects might not be uniquely determined by the observed data, even when strict parametric models are employed.
One recent approach to achieving identification in missing data analysis is to use auxiliary variables as shadow variables \citep{dHaultfoeuille2010,WangShaoKim2014,MiaoTchetgenTchetgen2018,BreunigHaan2021,MiaoLiuLiTchetgenTchetgenGeng2024}, which are required to be related to the missing variable while independent of the missingness process, conditional on both the observed data and missing variable. They are commonly used in survey sampling designs and appear in numerous empirical studies \citep{Kott2014}. In some instances, researchers impose certain structural assumptions on the distribution law, incorporating existing observed variables as shadow variables. For example, \cite{YangWangDing2019} proposed a valuable identification strategy for average treatment effects in the presence of outcome-independent missing covariates, leveraging the outcome as a suitable shadow variable.\label{yang:identification}
Although existing shadow variable-based methods were not explicitly developed for causal mediation analysis, they can, in principle, be extended to this setting. However, these methods typically restrict the shadow variable to originate from either externally collected auxiliary variables or existing observed variables satisfying certain independence conditions, but not both.

Once identifiability is established, another prominent challenge concerns estimation and inference, primarily arising from the estimation process of the missingness mechanism. It typically relies on solving a Fredholm integral equation of the first kind \citep{Kress_LinearIntegralEquations}, which is  known to be an ill-posed inverse problem, leading to notably slow convergence of the estimator toward its true value \citep{NeweyPowell2003}.
For nonignorable missing outcome data analysis, \cite{MiaoLiuLiTchetgenTchetgenGeng2024} tackled the ill-posedness problem by imposing a specific parametric model for an odds ratio function, thereby imposing certain structural constraints on the underlying model. They further developed the semiparametric efficiency bound under this modeling framework.
For causal effect estimation with nonignorable missing covariates, doubly robust estimators have been proposed \citep{SunLiu2021}, but they also rely on a correctly specified model of the odds ratio function.
However, bias may arise due to model specification errors, which are more likely to occur in the presence of missing values.
\cite{YangWangDing2019} proposed a nonparametric two-stage least squares approach for the average treatment effect to avoid model misspecification and proved its consistency. While their primary focus is on treatment effect identification, 
our work instead investigates the asymptotic properties and efficient inference that they did not explore.
The causal inference literature still lacks a well-developed asymptotic theory for nonparametric estimation under nonignorable missing covariates,  and further efforts are required to develop feasible nonparametric inference approaches.\label{yang:nonparametric}

\subsection{Our contributions}\label{subsec:contribution}
This paper presents a nonparametric methodology for identifying, estimating, and inferring the mediation effects in the presence of nonignorable missing confounders. Our main contributions are as follows. 

First, we propose a general shadow variable framework for identification, which enables the selection of shadow variables from either existing observed variables that satisfy specific independence conditions or externally collected auxiliary data. Within this flexible framework, we show that the mediation effect of interest can be expressed as a weighted average of an iterated conditional expectation. The weight is given by an inverse nonresponse probability, as specified by an integral equation. At the same time, the nuisances in the iterated conditional expectations are represented as some ratios of observed conditional expectations.\label{contribution:first} 

Second, based on the identification results, we propose a sieve-based iterative outward (SIO) estimation approach that starts from the innermost and iterates outward.  Our approach eliminates the need to parametrize the missingness mechanism, the conditional densities of the exposure and mediator, and the conditional expectation of the outcome. To improve robustness and finite-sample performance, we employ a one-step weighted least-squares procedure to sequentially estimate each ratio-structured nuisance function during the construction of the SIO estimator.

Third, we establish the semiparametric efficiency theory for the proposed estimator, which is technically challenging because the nuisances are intertwined and ill-posedness accumulates, influencing all subsequent estimates. While \citet{MiaoLiuLiTchetgenTchetgenGeng2024} also studied semiparametric efficiency for nonignorable missing data, their analysis focuses on settings with a single missing outcome and relies on a parametric specification of the odds ratio function. In contrast, we address the more complex problem of multivariate missing covariates under a fully nonparametric model, for which existing techniques and results are not directly applicable. To address these challenges, we first introduce a weaker pseudo-metric to achieve the requisite convergence rate for establishing asymptotic normality, allowing the weight estimator to converge to its true value faster than $n^{-1/4}$ under this metric. We then introduce a representer to capture the cumulative influence in the iterative outward estimation process, derive the influence function of the parameter of interest, and establish its asymptotic normality.
We show that the proposed estimator is locally efficient,  attaining the semiparametric efficiency bound under our shadow variable model.
By comparing this bound with the one established by \cite{TchetgenTchetgenShpitser2012} for complete data, we quantify the efficiency loss due to missing data. It offers heuristic insights into the impact of missingness on causal mediation analysis. Moreover, 
we propose a cross-fitted debiased machine learning approach based on the efficient influence function, which accommodates flexible estimation methods that excel in high-dimensional settings while enabling valid statistical inference.

\subsection{Organization}
The remainder of the paper is structured as follows. \Cref{sec:mediation_and_identification} presents the identification formulas of mediation functionals under our general shadow variable framework. \Cref{sec:estimation_asym} proposes a nonparametric estimation approach.
\Cref{sec:large_sample_theory} develops the large-sample theory, including results on consistency and convergence rates, asymptotic normality, and semiparametric efficiency, and also introduces a cross-fitted debiased machine learning approach.
\Cref{sec:simulation} presents simulation studies, and \Cref{sec:application} provides a real data example. 
\Cref{sec:discussion} concludes the paper.
We have included all technical details, additional simulation studies, and further elaborations in the supplementary material.
 
Throughout, we let $f$ denote the probability density or mass function. We use capital letters for random variables and lowercase letters for their realizations. The calligraphic letter $\calV$ denotes the support of $V$. For a function $g$,
define $\|g\|_2= \{\int_\calV \Abs{g(v)}^2f(v)dv \}^{1/2}$ to be its $\calL_2$ norm, and $\|g\|_\infty=\sup_{v\in\calV}\Abs{g(v)}$ to be its supremum norm. We write $\calL_2(V)=\{g:\calV\mapsto\mR:~\|g\|_2<\infty\}$. We also regard a vector of random variables as a set and use set operations, for example, $(V_1,V_2)\cap (V_1,V_3)=V_1$ and $(V_1,V_2)\backslash (V_1,V_3)=V_2$.
A list of notation is summarized in Supplementary \Cref{sec:app_notations_and_assums} for ease of reference.

\section{Mediation functionals and identification strategy}\label{sec:mediation_and_identification}

\subsection{Basic framework of mediation analysis}\label{ssec:def_mediation}
We first introduce the basic framework of mediation analysis in the complete-data case.
Suppose $T$ is a binary treatment variable, $Y$ an outcome of interest, and $M$ a (possibly vector-valued) mediator variable. 
Let $\X=(X_1,\ldots,X_p)$ be a vector of $p$-dimensional covariates.
We use the potential outcomes notation to define mediation effects. 
Let $Y(t,m)$ denote the potential outcome under treatment status $T=t$ and mediator $M=m$, and $M(t)$ represent the potential value of the mediator under treatment status $T=t$. Let $Y(t,M(t\pprime))$ denote the potential outcome for an individual, where we do not specify the actual level of $M$, but set it to what it would have been if treatment had been $T=t\pprime$. 
Following \cite{Pearl2001}, the natural indirect effect (NIE) is defined as $\xi^{(t)}\triangleq E\{Y(t,M(1))-Y(t,M(0))\}$, and the natural direct effect (NDE) is defined as $\zeta^{(t)}\triangleq E\{Y(1,M(t))-Y(0,M(t))\}$ for $t\in\{0,1\}$. Note that $Y(t)=Y(t,M(t))$. The total effect (or average treatment effect, ATE), defined as $\tau\triangleq E\{Y(1)-Y(0)\}$, can thus be decomposed as $\tau=\xi^{(t)}+\zeta^{(1-t)}$. Intuitively, NIE quantifies the difference in mean potential outcomes when switching the potential mediator values while keeping the treatment fixed. It quantifies the extent to which the treatment influences the outcome via the mediators.
In contrast, NDE quantifies the difference in mean potential outcomes when switching the treatment while keeping the potential mediator as its natural value at a given treatment assignment. It blocks the causal mechanism via the mediator and captures the direct influence of the treatment on the outcome. We impose the consistency assumption that $Y=Y(T,M(T))$ and  $M=M(T)$.
The following assumptions are standard in the literature and sufficient for identifiability of mediation effects \citep{ImaiKeeleYamamoto2010}.

\begin{assum}[Sequential ignorability]\label{ass:ignorability}
  For each $m\in\calM$ and $t,t\pprime\in\{0,1\}$, we assume that \\ $\{Y(t,m),M(t\pprime)\}\indep T\mid \X$ and $Y(t\pprime,m)\indep M(t)\mid T=t, \X$.
\end{assum}

\begin{assum}[Positivity]\label{ass:positivity}
  For any $m\in\calM$, $\x\in\calX$ and $t\in\{0,1\}$, we assume that $f(t\mid \x)>0$ and $f(m\mid t,\x)>0$.
\end{assum}

\Cref{ass:ignorability} assumes that the treatment is ignorable given the pre-treatment covariates, and the mediator is ignorable given the observed value of the treatment as well as the covariates. Under \Cref{ass:ignorability,ass:positivity}, \cite{ImaiKeeleYamamoto2010} showed that for $t,t\pprime\in\{0,1\}$, 
\begin{align}
\label{eq:imai}
E\left\{Y(t,M(t\pprime))\right\}=\iint_{\calM\times\calX}  E(Y \mid t,m,\x) f(m \mid t\pprime,\x) f(\x) dmdx.
\end{align}
The quantity $E\{Y(t, M(t\pprime))\}$ is referred to as a mediation functional. Although it has a closed-form expression as given on the right-hand side of \eqref{eq:imai}, it is generally not identifiable from the observed data when covariates $\X$ contain missing values.
Given the structural similarity among these functionals, we  focus on the fundamental quantity $\theta\triangleq E\{Y(1,M(0))\}$ and later extend the results to causal mediation effects.
To simplify notations, let $\gamma_0(m,\x)\triangleq E(Y\mid T=1,m,\x)$ and $\eta_0(\x)\triangleq E\{\gamma_0(M,\x)\mid T=0,\x\}$.
Then \eqref{eq:imai} indicates that $\theta=E\{\eta_0(\X)\}$. 

\subsection{Stratified identification}\label{ssec:identification}

In this section, we consider the scenario in which some components of $\X$ have missing values.
We provide a general shadow variable framework for identification,  which can be selected from either existing observed variables or externally collected auxiliary data. Within this framework, we develop a stratified identification strategy. The key idea is to partition the population into overlapping strata, with each stratum corresponding to a specific missingness pattern (excluding the complete-case pattern). Each stratum includes individuals from two groups: those with complete data and those exhibiting the specified missingness pattern. Within a given stratum, specific covariates that are only partially observed in the overall population may become fully observed. This enables us to use the available observed information efficiently for identification.

\begin{table}
  \renewcommand{\arraystretch}{1.1} 
  \caption{An example where $\X=(X_1,X_2,X_3)$, and $\X_\mis=(X_1,X_2)$.}
  \label{table:example_notation}
  \centering
  \begin{tabular}{c|cccccc}
  \hline%
  Patterns & $X_1$ & $X_2$ & $X_3$ & $\R$ & $\bar{R}$ & $\X_{\R}$ \\ \hline
  1 & $\checkmark$ & $\checkmark$ & $\checkmark$ & (1,1,1) & 1 & $(X_1,X_2,X_3)$ \\
  2 & $\checkmark$ & ? & $\checkmark$ & (1,0,1) & 0 & $(X_1,X_3)$ \\
  3 & ? & $\checkmark$ & $\checkmark$ & (0,1,1) & 0  & $(X_2,X_3)$ \\
  4 & ? & ? & $\checkmark$ & (0,0,1) & 0  & $X_3$ \\
  \hline
  \end{tabular}
\end{table}

Let $\X_\mis$ denote the components of covariates that are not fully observed.
Let $\R=(R_1,\ldots,R_p)$ be a vector of missing indicators such that $R_j=1$ if $X_j$ is observed and zero otherwise. Let $\calR=\{\r\in\{0,1\}^p:f(\r)>0\}$ be the set of all missingness patterns. In particular, let $\1$ denote the complete-case pattern $(1,1,\ldots,1)$.  Let $\X_{\r}=\{X_j:r_j=1\}$ be the observed portion of $\X$ for a pattern $\r\in\calR$. Therefore, we observe a pair $(\R,\X_{\R})$ for each individual. Denote $\bar{R}=\prod_{j=1}^pR_j$ indicating whether all covariates are observed. \Cref{table:example_notation} is an illustrative example with $p=3$ and $|\calR|=4$. 
Note that the joint distribution $f(\r,t,m,y,\x)$ is identifiable if and only if the missingness mechanism $f(\r\mid t,m,y,\x)$ is identifiable because $f(\r,t,m,y,\x)=f(\r\mid t,m,y,\x)f(\R=\1,t,m,y,\x)/f(\R=\1\mid t,m,y,\x)$.
Let $\fX\subset (T,M,Y,\X)$ be random variables that sufficiently capture the information about the missing mechanism such that $f(\r\mid t,m,y,\x)\equiv f(\r\mid \fx)$ holds for all $\r\in\calR$, where $\fx$ represents the realization of $\fX$. Knowledge of $\fX$ can help identify the underlying distribution as discussed later. In the absence of prior knowledge, we can roughly set $\fX=(T,M,Y,\X)$, indicating that the missingness of confounders may depend on themselves and on all other observed variables. 

The goal is to identify $f(\r\mid \fx)$ using observed data.
Noting that for each pattern $\r\in\calR\backslash\{\1\}$, $\X_{\r}$ are fully observable among the units with $\R\in\{\1,\r\}$. We therefore propose a stratified identification strategy to utilize this information efficiently. Specifically, we partition all units into $|\calR|-1$ overlapped strata, indexed by $\r\in\calR\backslash\{\1\}$, where the stratum $\r$ contains units with $\R\in\{\1,\r\}$.
Let $f_{\r}(\cdot)\triangleq f(\cdot\mid \R\in\{\1,\r\})$ be the density restricted on the subpopulation within the stratum $\r$, and $E_{\r}(\cdot)$ denote the expectation induced by $f_{\r}$. Then $f_{\r}(\R=\1\mid \fx)$ indicates the conditional probability of units without missing values in the stratum $\r$. Denote $\delta_{\r}(\fx)={f_{\r}(\R=\1\mid \fx)}^{-1}$, and $\omega_0(\fx)={f(\R=\1\mid \fx)}^{-1}$. We show in Supplementary \Cref{ssec:app_ident_distribution} that
$\omega_0(\fx) =  1+\textstyle\sum_{\r\in\calR\backslash\{\1\}} \{\delta_{\r}(\fx)-1\},$
and $f(\r\mid \fx)=\omega_0(\fx)^{-1}\{\delta_{\r}(\fx)-1\}$ for $\r\in\calR\backslash\{\1\}$.
Therefore, once $\delta_{\r}(\fx)$ is identified for each $\r\in\calR\backslash\{\1\}$, the missing mechanism  $f(\r\mid \fx)$ is identified. 
We adopt the shadow variable framework to identify $\delta_{\r}(\fx)$ using observed data in stratum $\r$. Suppose that the shadow variables $\Z$ are fully observed and selected based on 
$\fX$, satisfying the following assumption.

\begin{assum}[Shadow variables]\label{ass:shadow}
	 $\Z\indep \R\mid \fX$.
\end{assum}

\begin{figure}[h!]
  \centering
  \begin{subfigure}{.45\linewidth}
    \centering
    \includegraphics[width=.6\textwidth]{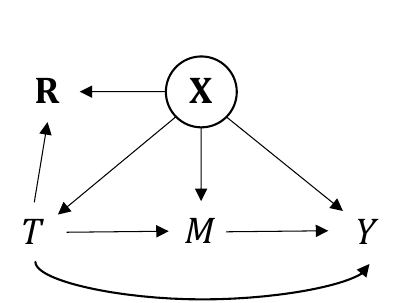}
    \caption{ $\fX=(T,\X)$ and $\Z=(M,Y)$. }
  \end{subfigure}
  \begin{subfigure}{.45\linewidth}
    \centering
    \includegraphics[width=.6\textwidth]{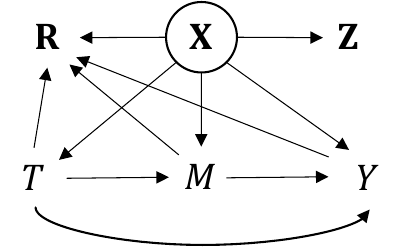}
    \caption{$\fX=(T,M,Y,\X)$ and an auxiliary $\Z$.}
  \end{subfigure}
  \caption{Causal directed acyclic graphs illustrating \Cref{exa:mo_independent,exa:auxiliary}.}
  \label{fig:DAG_example}
  \vspace{-5mm}
\end{figure}

\Cref{ass:shadow}  requires that shadow variables do not directly impact the missingness. In other words, once $\fX$ are present, the shadow variables $\Z$ will not provide additional information on the missing mechanism. 
\Cref{ass:shadow} implies that observed variables not included in $\fX$ can be considered as potential candidates for $\Z$. Our framework accommodates two common types of shadow variables. The first type involves using existing fully observed variables that do not influence the missing mechanism. For example, \citet{YangWangDing2019} introduced an outcome-independent missingness assumption that allows the outcome to serve as a shadow variable for identifying causal effects when confounders are subject to nonignorable missingness. \cite{ZuoGhoshDingYang2024} applied similar strategies to identify mediation effects when the mediator, the outcome, or both are missing. Related approaches have also been proposed in the literature on nonignorable missing outcomes, in which specific covariates are assumed to be excluded from the missingness mechanism and used as shadow variables; see, for example, \citet{WangShaoKim2014} and \citet{ZhaoMa2022}. The second type involves externally collected  auxiliary variables that serve as shadow variables, such as surrogates or error-prone proxies for variables subject to missingness \citep{dHaultfoeuille2010,MiaoTchetgenTchetgen2016,BreunigHaan2021,MiaoLiuLiTchetgenTchetgenGeng2024}. We provide further illustration through the following examples and \Cref{fig:DAG_example}.

\begin{example}[Mediator/outcome-independent missingness]\label{exa:mo_independent}
 In longitudinal studies where covariates are measured long before  the mediator and outcome occur, it is reasonable to assume that the mediator and outcome do not affect the missingness of covariates. In such cases, the mediator and outcome can be used as shadow variables, with $\fX=(T,\X)$ and $\Z=(M,Y)$ satisfying \Cref{ass:shadow}.
\end{example}

\begin{example}[Auxiliary variables]\label{exa:auxiliary}
  If there is no prior knowledge of the missingness mechanism, then $\fX=(T,M,Y,\X)$, and auxiliary variables such as proxies or  mismeasured versions of the missing  covariates are often used as shadow variables. For example, in a study evaluating children's mental health in Connecticut \citep{ZahnerPawelkiewiczDeFrancescoAdnopoz1992}, the teacher's assessment was prone to missingness, and, as \cite{IbrahimLipsitzHorton2001} suggested, a separate evaluation by parents served as a valid shadow variable.
\end{example}

\label{exam:shadow}
To see how shadow variables help identify $\delta_\r$ based on observed data in the stratum $\r\in\calR\backslash\{\1\}$, note that by definition, $\delta_\r$ satisfies the moment condition $E_\r\{\bar{R}\delta_\r(\fX)\mid \fX\}=1$. However, this equation is infeasible because the conditioning set $\fX$ may include unobserved components.
To address this, we first leverage shadow variables to obtain the equivalent condition $E_\r\{\bar{R}\delta_\r(\fX)\mid \fX,\Z\}=1$ under \Cref{ass:shadow}. We then integrate out the unobserved components of $\fX$, $\fX\cap\X_{-\r}$, where $\X_{-\r}=\X\backslash\X_{\r}$, resulting in the feasible moment condition $ E_\r\{\bar{R}\delta_\r(\fX)\mid \fZ_\r\}=1$. Here, $\fZ_\r=\fX\cup\Z\backslash\X_{-\r}$ denotes the set of variables observable in stratum $\r$. Let $\calA_{\r}:\calL_2(\fX)\rightarrow\calL_2(\fZ_{\r})$ be the linear operator defined by $\calA_{\r}(\delta)\triangleq E_{\r}\{\bar{R}\delta(\fX)\mid \fZ_{\r}\}$. Then the feasible moment condition can be written as $\calA_{\r}(\delta_{\r})=1$. 
Let $\fZ=\bigcap_{\r\in\calR\backslash\{\1\}}\fZ_\r=\fX\cup\Z\backslash\X_\mis$ denote the set of variables observed in all strata.
We use \Cref{exa:mo_independent} to illustrate the notation further.

\begin{exctd}[\ref*{exa:mo_independent}]\label{exa:mo_independent_contd}
  Considering the population in \Cref{table:example_notation} and \Cref{exa:mo_independent}, we have $\fX=(T,X_1,X_2,X_3)$, and $\fZ=(T,M,Y,X_3)$. 
  For the stratum $\r=(1,0,1)$, we have $\fZ_{\r}=(T,M,Y,X_1,X_3)$, and $\delta_{\r}(\fx)=f(\R=\1\mid \R\in\{\1,\r\},t,x_1,x_2,x_3)^{-1}$ satisfies the 
  integral
  equation
  \begin{align*}
  E\left[\frac{\bar{R}}{f(\R=\1\mid \R\in\{\1,\r\},T,X_1,X_2,X_3)}\mid \R\in\{\1,\r\},T,M,Y,X_1,X_3\right]=1.
  \end{align*}
  We emphasize that although $X_1$ has missing values in the overall sample, it is fully observed in the stratum $\r=(1,0,1)$, and thus can be leveraged to identify $\delta_{\r}$.
  The same applies to $\r=(0,1,1)$ and $\r=(0,0,1)$.
\end{exctd}

To ensure a unique solution to the equation $\calA_{\r}(\delta_{\r}) = 1$ for each stratum $\r$, we impose the following completeness assumption.
\begin{assum}[Completeness]\label{ass:completeness}
	 ${E\{g(\fX)\mid \fZ\}=0}$ implies $g(\fX)=0$ for any square-integrable function $g$.
\end{assum}

Note that $\fX\backslash\fZ=\X_\mis\cap\fX$ and $\fZ\backslash\fX=\Z$.
\Cref{ass:completeness} can be intuitively interpreted as requiring the shadow variable $\Z$ to be correlated with and exhibit sufficient variation relative to {the components of $\X_\mis$ that affects the missingness mechanism}. For instance, in the case of categorical $\Z$ and $\X_\mis$, it requires $\Z$  having at least as many categories as $\X_\mis\cap\fX$.
Completeness is a fundamental concept in statistics and is widely used for nonparametric identification in literature \citep{NeweyPowell2003,DHaultfoeuille2011}. It holds for a variety of parametric or semiparametric models, such as exponential families. We provide additional discussions about this condition in Supplementary \Cref{ssec:app_discuss_completeness}.
To see how \Cref{ass:completeness} ensures the identifiability of $\delta_\r$, let $\calA:\calL_2(\fX)\rightarrow\calL_2(\fZ)$ denote the linear operator defined by $\calA(\delta)\triangleq E\{\bar{R}\delta(\fX)\mid \fZ\}$.
As demonstrated in Supplementary \Cref{ssec:app_strengthened_moment}, the equation $\calA_{\r}(\delta_{\r})=1$ implies the weaker conditional moment restriction $\calA(\delta_{\r})=f(\R\in\{\1,\r\}\mid\fZ)$. The completeness \cref{ass:completeness} guarantees the injectivity of $\calA$, thereby ensures the identification of $\delta_{\r}$. 
We also assume that $f(\R=\1\mid \fx)$ is bounded away from 0 to rule out degeneracy of the missingness mechanism. 
We summarize the above procedure and establish identifications of $\gamma_0$, $\eta_0$ and $\theta$  in the following theorem.

\begin{theorem}\label{thm:iden_formula}
  Suppose \Crefrange{ass:ignorability}{ass:completeness} hold. Then, $\delta_{\r}(\fx)$ is identified for $\r\in\calR\backslash\{1\}$, and hence, $\omega_0(\fx)=1+\sum_{\r\in\calR\backslash\{\1\}}\{\delta_{\r}(\fx)-1\}$ is also identified. 
  In addition, we have:
  \begin{enumerate}[(i)]
    \item  $\gamma_0(m,\x)$ is identified as the following ratio of observed conditional expectations:
    \begin{align}\label{eq:iden_gamma}
    \gamma_0(m,\x) = \frac{E\{Y\omega_0(\fX)\mid \R=\1,T=1,m,\x\}}{E\{\omega_0(\fX)\mid \R=\1,T=1,m,\x\}}.
    \end{align}
    \item $\eta_0(\x)$ is identified as the following ratio of observed conditional expectations:
    \begin{align}\label{eq:iden_eta}
    \eta_0(\x) = \frac{E\{\gamma_0(M,\X)\omega_0(\fX)\mid \R=\1,T=0,\x\}}{E\{\omega_0(\fX)\mid \R=\1,T=0,\x\}}.
    \end{align}
    \item
    The mediation functional $\theta$ is identified as the following weighted expectation:
      \begin{align} \label{eq:iden_theta}
      \theta =  E\{\bar{R}\omega_0(\fX)\eta_0(\X)\}.
      \end{align}
  \end{enumerate}
\end{theorem}

\begin{remark}\label{remark:mar}
  Our framework covers the missing-at-random mechanism, corresponding to $\fX\cap\X_\mis=\emptyset$, as a special case, where a trivial choice of $\Z=\emptyset$ satisfies \Cref{ass:shadow,ass:completeness}. In other words, \Cref{thm:iden_formula} remains applicable for identifying $\theta$ without the need of shadow variables. We elaborate on this case in Supplementary \Cref{sec:supp_mar}.
\end{remark}

\section{Nonparametric estimation}\label{sec:estimation_asym}
In this section, we consider the estimation of $\theta$. Suppose we observe a set of $n$ independent and identically distributed realizations of observed variables $\O=(\R,T,M,Y,\X_{\R},\Z)$. 
We note that by specifying working models $f(\r\mid \fx;\mathbf{a}_1)$, $f(y\mid \R=\1,t,m,\x;\mathbf{a}_2)$ and $f(m\mid \R=\1,t,\x;\mathbf{a}_3)$, one can construct a straightforward parametric estimator of $\theta$ based on the identification formula in \Cref{thm:iden_formula}, analogous to the approach used in the complete-data setting of \citet{TchetgenTchetgenShpitser2012}.
However, this parametric strategy has two main limitations. First, it is vulnerable to model misspecification, which may lead to biased estimates. Second, the ratio structures inherent in $\gamma_0$ and $\eta_0$ may introduce numerical instability.
While \citet{HuangHuangLintonZhang2024} proposed a nonparametric method for mediation analysis using inverse probability weighting with nonparametrically estimated calibration weights, their approach does not address the complications introduced by nonignorable missing covariates. In contrast, our study focuses on this more challenging setting, requiring fundamentally new methodological developments. \label{rev:huang-compare1}

Below, we propose a nonparametric one-step approach that directly models and estimates nuisance functions sequentially. 
We begin with the nonparametric estimation of $\delta_{\r}$ for each $\r\in\calR\backslash\{\1\}$ from the integral equation $\calA_{\r}(\delta_{\r})=1$. 
We estimate the inverse of the missingness mechanism, $\delta_{\r}$, directly rather than $f_{\r}(\R=\1\mid \fx)$ to avoid instability arising from inverse probability weights in subsequent estimations.
Heuristically, the integral equation implies that $\delta_{\r}$ solves $\inf_{\delta\in\Delta} E_{\r}\{\calA_{\r}(\delta)-1\}^2$, where $\Delta\subset \calL_2(\fX)$ is the function space of $\delta_{\r}$.
Since the objective function depends only on the observed data in stratum $\r$, this suggests estimating $\delta_{\r}$ by solving its empirical analogue using samples from stratum $\r$. However, it encounters two obstacles. First, the operator $\calA_{\r}$, equivalently the conditional expectation $E_{\r}(\cdot\mid \fZ_{\r})$, is unknown. Second, without imposing parametric assumptions on the function form of $\delta_{\r}$, its value space $\Delta$ is an infinite-dimensional set; thus, it is impractical to derive an estimator of $\delta_{\r}$ over $\Delta$ with finite samples. 

We address the first problem by estimating the operator $\calA_{\r}$ via series least squares estimation. 
Let $\fz_{\r}$ denote the realization of $\fZ_{\r}$.
For a specific pattern $\r\in\calR\backslash\{\1\}$, let $\{p_j(\cdot)\}_{j=1}^\infty$ denote a sequence of known basis functions (such as power series, splines, Fourier series, etc.), with the property that its linear combination can approximate any square-integrable real-valued function of $\fz_{\r}$ well.
Let $\p^{k_n}(\fz_{\r})= (p_1(\fz_{\r}),\ldots,p_{k_n}(\fz_{\r}))\trans$, and $\mathbf{P}$ be the $n$-by-$k_n$ random matrix whose $i$-th row is $I(\R_i\in\{\1,\r\})\p^{k_n}(\fZ_{\r,i})\trans$.
{We use universal notations $p_j(\cdot)$, $k_n$, and $\mathbf{P}$ for different $\r\in\calR\backslash\{1\}$ for simplicity, even though they may vary with $\r$, whenever no confusion arises.}
For any random variable $V$ with $n$ realizations $\{V_i\}_{i=1}^n$, the series estimator is given by 
\begin{align}\label{eq:series_LS}
\wh{E}_{\r}(V\mid \fz_{\r}) = \sum_{i=1}^n I(\R_i\in\{\1,\r\})V_i \p^{k_n}(\fZ_{\r,i})\trans (\mathbf{P}\trans\mathbf{P})^{-1}\p^{k_n}(\fz_{\r}).
\end{align}
Then we estimate the operator $\calA_{\r}(\cdot)$ by
$\wh\calA_{\r}(\delta) = \wh{E}_{\r}(\bar{R}\delta\mid \fZ_{\r})$ for any $\delta\in\Delta$.
To deal with the second problem, we use a sieve space $\Delta_{J_n}$ to approximate $\Delta$, which is typically finite-dimensional and becomes dense in $\Delta$ as $n$ increases. An estimator $\wh\delta_{\r}$ of $\delta_{\r}$ is then obtained by solving  
  \begin{align}
  \label{eq:criterion_empirical_delta}
  \min_{\delta\in\Delta_{J_n}} \frac{1}{n}\sum_{i=1}^n I\big(\R_i\in\{\1,\r\}\big)\big\{\wh\calA_{\r,i}(\delta)-1\big\}^2,
  \end{align}
where $\wh\calA_{\r,i}(\delta)=\wh{E}_{\r}(\bar{R}\delta\mid \fZ_{\r,i})$.
The construction of $\Delta_{J_n}$ requires careful consideration. Since $\delta_{\r}$ is the inverse of a probability function, the function space $\Delta$ should accommodate bounded functions with values greater than 1. The widely used linear sieve space \citep[see e.g.][]{Chen2007Handbook} may not be the best choice, as approximating a function that is always greater than 1 with a linear function may be unreasonable.  We suggest using a generalized linear sieve space. Let $\{q_j(\cdot)\}_{j=1}^\infty$ denote a sequence of known basis functions with the property that their linear combination can approximate any square-integrable real-valued function of $\fx$ well. Let $\rho(\cdot)$ be a link function that is sufficiently smooth and strictly monotone,  and possesses a range of values greater than 1. For example, one can use the inverse logistic function $\rho(v) = 1 + \exp(-v)$ or the double-exponential function $\rho(v) = \exp\{\exp(v)\}$. Then we construct the sieve space by 
  $\Delta_{J_n} = \{\delta:\delta(\fx)=\rho\{\sum_{j=1}^{J_n}\pi_jq_j(\fx)\},~\pi_1,\ldots,\pi_{J_n}\in\mR \}.$
Once we have obtained estimators $\wh\delta_{r}$ for each $\r\in\calR\backslash\{1\}$, we can finally estimate $\omega_0$ by $\wh\omega(\cdot)=1+\sum_{\r\in\calR\backslash\{\1\}}\{\wh\delta_{\r}(\cdot)-1\}$.

Next, we estimate  $\gamma_0(m,\x)$ and $\eta_0(\x)$ sequentially. With the estimator $\wh\omega$, one can proceed with a straightforward ``plug-in" procedure to derive estimators of $\gamma_0$ and $\eta_0$, by performing series least squares similar to \eqref{eq:series_LS} for the numerator and denominator of \eqref{eq:iden_gamma} and \eqref{eq:iden_eta} separately, and substituting the estimated values of unknown functions into the pseudo-outcomes. However, estimators of this type may be unstable due to their ratio form. We instead introduce a one-step estimation approach to pursue stability. To utilize this, we notice that $\gamma_0\in\Gamma\subset\calL_2(M,\X)$ in \eqref{eq:iden_gamma} is the unique solution to the criterion function of the weighted $\calL_2$-projection problem
$
\inf_{\gamma\in \Gamma} E\left[T\bar{R}\omega_0(\fX)\{Y-\gamma(M,\X)\}^2\right],
$
and $\eta_0\in\Lambda\subset\calL_2(\X)$ in \eqref{eq:iden_eta} is the unique solution to 
$
\inf_{\eta\in \Lambda} E\left[(1-T)\bar{R}\omega_0(\fX)\{\gamma_0(M,\X)-\eta(\X)\}^2\right].
$
Therefore, it suggests estimators $\wh\gamma$ and $\wh\eta$ that solve the empirical optimization problems over corresponding sieve spaces.
Specifically, let $\{u_j(\cdot)\}_{j=1}^\infty$ (resp. $\{v_j(\cdot)\}_{j=1}^\infty$) denote a sequence of known basis functions with the property that its linear combination can approximate any square-integrable real-valued function of $(m,\x)$ (resp. $\x$) well.
Then $\Gamma_{l_n} = \{\gamma:\gamma(m,\x)=\sum_{j=1}^{l_n}\pi_ju_j(m,\x),~\pi_1,\ldots,\pi_{l_n}\in\mR\}$ and $\Lambda_{s_n} = \{\eta: \eta(\x)=\sum_{j=1}^{s_n}\pi_jv_j(\x),~\pi_1,\ldots,\pi_{s_n}\in\mR\}$, 
with $l_n\rightarrow\infty$ and $s_n\rightarrow\infty$ slowly as $n\rightarrow\infty$, are finite-dimensional linear sieves for $\Gamma$ and $\Lambda$, respectively.
We then obtain  $\wh\gamma$ by solving
\begin{align}
\label{eq:criterion_empirical_gamma}
\min_{\gamma\in\Gamma_{l_n}} \frac{1}{n}\sum_{i=1}^{n} T_i\bar{R}_i\wh\omega(\fX_i)\{Y_i-\gamma(M_i,\X_i)\}^2,
\end{align}
and further  $\wh\eta$ by referring to $\wh\gamma$ as the new pseudo-outcome and solving
\begin{align}
\label{eq:criterion_empirical_eta}
\min_{\eta\in\Lambda_{s_n}} \frac{1}{n}\sum_{i=1}^{n} (1-T_i)\bar{R}_i\wh\omega(\fX_i)\{\wh\gamma(M_i,\X_i)-\eta(\X_i)\}^2.
\end{align}
Notably, the first-order conditions of \eqref{eq:criterion_empirical_gamma} and \eqref{eq:criterion_empirical_eta} provide closed-form expressions for $\wh\gamma$ and $\wh\eta$ as shown in Supplementary \Cref{ssec:pf_thm:convergence_rate_gamma_eta}.
An estimator of $\theta$ is finally obtained based on \Cref{thm:iden_formula} using the empirical weighted average over the observed samples:
\begin{align}
\label{eq:theta_hat}
\wh\theta = \frac{1}{n}\sum_{i=1}^{n} \bar{R}_i\wh\omega(\fX_i)\wh\eta(\X_i).
\end{align}

\begin{remark}\label{ssec:sel_tuning}
The selection of sieve dimensions $(J_n,k_n,l_n,s_n)$ is a critical issue for practical implementation.
Several data-driven methods for selecting tuning parameters in series estimation have been discussed in \cite{Li1987} and \citet[Section 15.2]{LiRacine2007}. Building on these results, we propose a data-driven procedure for selecting tuning parameters by minimizing penalized loss functions. We first consider the selection of $(J_n,k_n)$ when estimating $\delta_\r$ for $\r\in\calR\backslash\{\1\}$. Let $\wh\calA_\r^{k_n}$ and $\wh\delta_{\r}^{J_n,k_n}$ denote the estimators of $\calA_{\r}$ and $\delta_\r$, respectively, obtained using the basis $\p^{k_n}$ and the sieve space $\Delta_{J_n}$ in \eqref{eq:series_LS}-\eqref{eq:criterion_empirical_delta}. Define the penalized loss function $\ell_1(J_n,k_n)=\sum_{i=1}^n I(\R_i\in\{\1,\r\})\{\wh\calA_{\r,i}^{k_n}(\wh\delta_{\r}^{J_n,k_n})-1\}^2(1+2k_n/n)$. Recalling the regularity condition $k_n\ge J_n$ in \Cref{ass:supp_sieve_delta_exter} of the supplementary material, we consider a given $k_n(J_n)\ge J_n$, for example, $k_n(J_n)=J_n+1$. Then, we select $\wh J_n$ by minimizing $\ell_1(J_n,k_n(J_n))$ over a candidate set $J_n\in\mathbb{J}$, and select $\wh k_n=k_n(\wh J_n)$. A similar procedure can be applied to select $(l_n,s_n)$. For example, let $\wh\gamma^{l_n}$ denote the estimator of $\gamma$  using the sieve space $\Gamma_{l_n}$ in \eqref{eq:criterion_empirical_gamma}. Define $\ell_2(l_n)=\sum_{i=1}^{n} T_i\bar{R}_i\wh\omega(\fX_i)\{Y_i-\wh\gamma^{l_n}(M_i,\X_i)\}^2(1+2l_n/n)$. Then, $\wh l_n$ is chosen by minimizing $\ell_2(l_n)$ over a candidate set $l_n\in\mathbb{L}$.
\end{remark}

\begin{remark}\label{remark:no_missing_huang-comparison1.5}
  In the special case where there is no missing data, simply setting $\wh\omega\equiv 1$ makes the proposed estimator and its accompanying asymptotic theory remain valid. From this perspective, our method also provides a novel nonparametric approach to mediation analysis for fully observed data, serving as an alternative to existing parametric methods and to the nonparametric approach proposed by \cite{HuangHuangLintonZhang2024}.
\end{remark}

\begin{remark}\label{remark:curse_dim}
  The nonparametric approach is known to suffer from the curse of dimensionality, as the precision of approximating an unknown function class using a sieve space in general depends negatively on the dimension of arguments. Some literature suggests using parametric methods for this reason.  We note here that if one has oracle knowledge about the function classes ($\Delta,\Gamma,\Lambda$) falling within certain parametric classes, our approach remains valid by finding solutions to the optimization problems \eqref{eq:criterion_empirical_delta}--\eqref{eq:criterion_empirical_eta} within these pre-specified parametric classes. In this case, our method can be viewed as a one-step parametric approach that directly models the nuisance functions, in contrast to the indirect approach of modeling density functions, as in \cite{TchetgenTchetgenShpitser2012}, thereby achieving greater stability in practice.
\end{remark}

\section{Asymptotic theory and efficient inference}\label{sec:large_sample_theory}
\subsection{Consistency and convergence rate}\label{ssec:convergence_rate}
We now investigate large sample properties of the proposed estimators.
Of primary interest is the asymptotic distribution of $\wh\theta$ because it is the foundation for statistical inference such as hypothesis testing and constructing confidence intervals.
For this purpose, intermediate results regarding the convergence rates of nuisance estimators $\wh\delta_{\r}$, $\wh\gamma$ and $\wh\eta$ are required. It is well known that asymptotic normality typically requires the nuisance estimators to converge at a rate faster than $n^{-1/4}$ under certain metric.
However, it is challenging in this study because solving integral equations is known to be ill-posed. It means, for each $\r\in\calR\backslash\{\1\}$, the convergence rate of $\wh\delta_{\r}$ to $\delta_{\r}$ under $\|\cdot\|_\infty$ may be very slow, which creates challenges in establishing the asymptotic normality. 
To address this,  we first prove the consistency of each $\wh\delta_{\r}$ under the supremum norm $\|\cdot\|_\infty$.  Then, we obtain a faster convergence rate for each $\wh\delta_{\r}$ under a weaker pseudo-metric $\|\cdot\|_{w,\r}$ defined by
\begin{align}
\label{eq:weak_norm}
\|\delta\|_{w,\r}^2 \triangleq E_{\r}\{\calA_{\r}(\delta)^2\} = E_{\r}\left[\left\{E_{\r}\big(\bar{R}\delta(\fX)\mid \fZ_{\r}\big)\right\}^2\right],~~\delta\in\Delta.
\end{align}
It is clear that $\|\delta\|_{w,\r}\le c_1\|\delta\|_2\le c_2\|\delta\|_\infty$ for any $\delta\in\Delta$ {and some constants $c_1,c_2$}.  
Moreover, we establish the convergence rates for $(\wh\gamma,\wh\eta)$ under both the $\calL_2$ norm and the supremum norm. These convergence results play a crucial role in establishing the asymptotic normality of $\wh\theta$.

Some technical conditions are required to establish the rates of convergence, which is listed and discussed in detail in Supplementary \Cref{sec:app_notations_and_assums}. For instance, we require that the functions in $\Delta$ are sufficiently smooth and bounded, and $\Delta$ is compact under ${\|\cdot\|_\infty}$. This is a commonly imposed condition in the nonparametric and semiparametric literature to deal with the ill-posed problem arising from discontinuity of the solution to the integral equation. It ensures that the consistency of $\wh\delta_{\r}$ under $\|\cdot\|_\infty$ is not affected by the ill-posedness \citep{NeweyPowell2003}. Moreover, we impose some standard regularity conditions for use of sieve approximation. For example, we assume that the approximation error decays at a polynomial rate. This requirement is known to be satisfied for various kinds of sieve bases \citep{Chen2007Handbook}, and the polynomial rate typically depends on the smoothness of the approximand and the dimension of arguments. 
The convergence rates of nuisance estimators are established in the following theorem.
\begin{theorem}\label{thm:rate_nuisance}
  Suppose \Crefrange{ass:supp_dgp}{ass:supp_sieve_delta_exter} and \ref{ass:supp_sieve_strengthened_rate}(i) in Supplementary \Cref{sec:app_notations_and_assums} hold.
  \begin{enumerate}[(i)]
    \item For each $\r\in\calR\backslash\{\1\}$, there exists a sequence $0<\lambda_{1n,\r}=o(n^{-1/4})$ such that
      \begin{align*}
      \|\wh\delta_{\r}-\delta_{\r}\|_\infty=o_p(1)
      ~~\mbox{and}~~
      \|\wh\delta_{\r}-\delta_{\r}\|_{w,\r}= O_p(\lambda_{1n,\r}).
      \end{align*}
    \item Additionally suppose \Crefrange{ass:supp_sieve_gamma}{ass:supp_technique_gamma_eta} and \ref{ass:supp_sieve_strengthened_rate}(ii)-(iii) in Supplementary \Cref{sec:app_notations_and_assums}  hold. For $\star\in\{2,\infty\}$, there exist sequences $0<\lambda_{2n,\star}=o(n^{-1/4})$ and ${0<\lambda_{3n,\star}=o(n^{-1/4})}$ such that 
    \begin{align*}
      \|\wh\gamma-\gamma_0\|_{\star} =  O_p(\lambda_{2n,\star})
      ~~\mbox{and}~~
      \|\wh\eta-\eta_0\|_{\star} =  O_p(\lambda_{3n,\star}).
    \end{align*}
  \end{enumerate}
\end{theorem}
\Cref{thm:rate_nuisance}(i) demonstrates that, for each $\r\in\calR\backslash\{\1\}$, the estimator $\wh\delta_{\r}$ converges in probability to $\delta_{\r}$ under the supremum norm $\|\cdot\|_\infty$, and converges at a rate faster than $n^{-1/4}$ under the weaker pseudo-metric $\|\cdot\|_{w,\r}$. \Cref{thm:rate_nuisance}(ii) shows that $\wh\gamma$ and $\wh\eta$ converge at a rate faster than $n^{-1/4}$ in both $\calL_2$ and supremum norms. 
The explicit rates of convergence depend on the smoothness parameters $(J_n,k_n,l_n,s_n)$ and are presented in \Cref{thm:weak_rate_delta,thm:convergence_rate_gamma_eta} of the supplementary material.
Specifically, the convergence rate of $\wh\delta_{\r}$ under $\|\cdot\|_{w,\r}$ is controlled by the estimation error of the series estimator \eqref{eq:series_LS} for the conditional mean and the approximation error of the sieve space $\Delta_{J_n}$ for the parameter space $\Delta$.
In contrast, $\wh\gamma$ achieves the standard nonparametric convergence rate in both $\calL_2$ and supremum norm \citep{Newey1997},
while $\wh\eta$ also achieves the conventional nonparametric convergence but is additionally affected by the approximation error of $\wh\gamma$ due to the iterative estimation process.

\subsection{Asymptotic normality}\label{ssec:asym_normal}
In this section we utilize  previous results to establish asymptotic normality of $\wh\theta$. 
It is still difficult to derive a linear representation of $\wh\theta$ using conventional techniques such as Taylor expansion. 
For example, due to the complexity of $\Delta_{J_n}$ with regularization, we do not have a closed-form expression of $\wh\delta_{\r}$ for each $\r\in\calR\backslash\{\1\}$.
In addition, the nuisances are intertwined and the estimators $\wh\gamma$ and $\wh\eta$ involve $\wh\omega$ (i.e., all $\wh\delta_{\r}$) in their explicit expressions.  For these reasons, we introduce a representer to depict the cumulative influence of intertwined nuisances for $\theta$ and help establish the linear representation of $\wh\theta$.
For $\r\in\calR\backslash\{\1\}$, let $\overline\calH_{\r}$ be the closure of the linear span of $\Delta$ under the metric $\|\cdot\|_{w,\r}$. Then $(\overline\calH_{\r},\|\cdot\|_{w,\r})$ is a Hilbert space with inner product $\langle \delta_1,\delta_2  \rangle_{w,\r} \triangleq  E_{\r}\{\calA_{\r}(\delta_1)\calA_{\r}(\delta_2)\}$
for $\delta_1,\delta_2\in\overline\calH_{\r}$.
We define
  \begin{small}
    \begin{align}
      &b_1(\x) \triangleq f(T=0\mid \x)^{-1}, ~~
      b_2(m,\x) \triangleq f(m\mid T=0,\x)f(T=1\mid \x)^{-1}f(m\mid T=1,\x)^{-1}, 
      \nonumber
      \\
      &\kappa(t,m,y,\x;\gamma,\eta)  
       \triangleq   \eta(\x) + (1-t)b_1(\x)\left\{\gamma(m,\x)-\eta(\x)\right\}  + tb_2(m,\x) \left\{y-\gamma(m,\x)\right\}, \label{eq:def_r_IF}
    \end{align}
  \end{small}
for any $\gamma\in\Gamma$ and $\eta\in\Lambda$.  We impose the following assumption.

\begin{assum}[Representer]\label{ass:representer} 
  For each $\r\in\calR\backslash\{\1\}$, there exists $\varrho_{\r}\in\Delta$ such that
  $\langle \varrho_{\r},\delta \rangle_{w,\r}
   = E_{\r}\{ \bar{R}\kappa(T,M,Y,\X;\gamma_0,\eta_0)\delta(\fX) \}$
  holds for any $\delta\in\overline\calH_{\r}$.
\end{assum}

\Cref{ass:representer} essentially guarantees $\sqrt{n}$-convergence of $\wh\theta$ to its truth. Similar conditions are also imposed in nonparametric instrumental variable literature \citep{AiChen2003,Santos2011,LiMiaoTchetgenTchetgen2023}. Note that the linear functional $\delta\mapsto E_{\r}\{ \bar{R}\kappa(T,M,Y,\X;\gamma_0,\eta_0)\delta(\fX)\}$ is continuous under $\|\cdot\|_{w,\r}$. By the Riesz representation theorem, there exists a unique $\varrho_{\r}\in\overline\calH_{\r}$ that satisfies the equation in \Cref{ass:representer}.
This assumption essentially requires the Riesz representer to fall within $\Delta$, which is automatically guaranteed if $\Delta$ itself is a closed Hilbert space, i.e., $\Delta=\overline\calH_{\r}$.
However, if this is not the case, we can still relax \Cref{ass:representer} by allowing $\varrho_{\r} \in \overline\calH_{\r}$ but $\varrho_{\r} \notin \Delta$. In such cases, we further assume that there exists $\varrho_{n,\r} \in \Delta_{J_n}$ such that $\|\varrho_{\r}-\varrho_{n,\r}\|_{w,\r} = O(n^{-1/4})$ similar to \cite{AiChen2003}. This condition also ensures the validity of the theoretical analysis. 
Combining \Cref{thm:rate_nuisance} and \Cref{ass:representer}, we establish the asymptotic normality of $\wh\theta$ in the following theorem.
\begin{theorem}\label{thm:asy_normal} 
    Suppose \Cref{ass:representer} and \ref{ass:supp_dgp}--\ref{ass:supp_sieve_strengthened_rate} in Supplementary \Cref{sec:app_notations_and_assums} hold.  We have 
    $\sqrt{n}(\wh\theta-\theta) = n^{-1/2}\sum_{i=1}^{n} \psi(\O_i) + o_p(1)$ with
    \begin{align*}
        \psi(\O) =&~   
        \bar{R}\omega_0(\fX) \kappa(T,M,Y,\X;\gamma_0,\eta_0)  +  \sum_{\r\in\calR\backslash\{\1\}} \calA_{\r}(\varrho_{\r}) \left[I(\R\in\{\1,\r\})-\bar{R}\delta_{\r}(\fX)\right] - \theta.
    \end{align*}
    It follows that $\sqrt{n}(\wh\theta-\theta)\rightarrow N(0,\sigma^2)$ in distribution, with $\sigma^2 =E\left\{\psi(\O)^2\right\}$.
\end{theorem}

We further provide a consistent estimator of $\sigma^2$ in Supplementary Appendix \ref{sec:CI}, which, together with Theorem \ref{thm:asy_normal}, enables valid confidence interval construction for $\theta$. Moreover, in Supplementary Appendix \ref{ssec:app_inference_NIE}, we extend the analytical framework to allow inference on the NIE, NDE, and ATE.

\begin{remark}\label{remark:EIF_no_missing}
    If no missingness occurs, that is, $\calR=\{\1\}$, $\bar{R}\equiv 1$, and $\omega_0\equiv 1$, the influence function of $\wh\theta$ turns out to be $\psi_\full(T,M,Y,\X) = \kappa(T,M,Y,\X;\gamma_0,\eta_0) - \theta$, which is the efficient influence function of $\theta$ for complete data given by \cite{TchetgenTchetgenShpitser2012}. We then let $\sigma^2_\full = E\{\psi_\full(T,M,Y,\X)^2\}$ denote the semiparametric efficiency bound for $\theta$ when $\X$ are fully observed. In the following section, we compare it with our scenario in which $\X$ contains missing values.
\end{remark}

\subsection{Efficiency}\label{ssec:efficiency}
\Cref{thm:asy_normal} presents the influence function for $\wh\theta$. Under certain regularity conditions, we demonstrate that $\wh\theta$ is locally efficient in the sense that it attains the semiparametric efficiency bound for $\theta$ in the semiparametric model $\calM_{sp}$, which imposes no restrictions on the observed data distribution other than the restriction that $\calA(\delta_{\r})=f(\R\in\{\1,\r\}\mid\fZ)$ holds for $\r\in\calR\backslash\{\1\}$. This specification is more flexible than alternatives such as $\mathcal{A}_{\r}(\delta_{\r}) = 1$ or Assumption~\ref{ass:shadow}, while still ensuring identification under the shadow variable framework.

\begin{assum}[Completeness]\label{ass:completeness_inverse}
  ${E\{g(\fZ)\mid \fX\}=0}$ implies $g(\fZ)=0$ for any squared-integrable function $g$.
\end{assum}

\Cref{ass:completeness_inverse} is a symmetric counterpart of \Cref{ass:completeness}. As previously discussed, \Cref{ass:completeness} guarantees the injectivity of the conditional mean operator $\calA$, while \Cref{ass:completeness_inverse} here ensures the injectivity of its adjoint operator $\calA\pprime$. This condition implies that the image space of $\calA$ contains any square-integrable real-valued function, which is a sufficient condition for the tangent space to incorporate the influence function.

\begin{theorem}\label{thm:EIF}
	The estimator $\wh\theta$ attains the semiparametric efficiency bound of $\theta$ in $\calM_{sp}$ at the submodel where \Cref{ass:completeness_inverse} holds.
\end{theorem}

\Cref{thm:EIF} reveals that $\sigma^2$ in \Cref{thm:asy_normal} is the semiparametric efficiency bound for $\theta$ when $\X$ are nonignorably missing, and shadow variables $\Z$ are available. This result allows us to quantify the information loss in estimating $\theta$ caused by the missingness in $\X$. We use a simplified notation $\kappa\triangleq\kappa(T,M,Y,\X;\gamma_0,\eta_0)$. We show in Supplementary \Cref{ssec:app_decomposition} that the influence function of $\wh\theta$ can be decomposed into two uncorrelated terms: 
\begin{align*}
  \psi(\O) = \psi_\full(T,M,Y,\X) + \sum_{\r\in\calR\backslash\{\1\}}\{ \calA_{\r}(\varrho_{\r})-\kappa\} \left[I(\R\in\{\1,\r\})-\bar{R}\delta_{\r}(\fX) \right],
\end{align*}
where the second term, denoted by $\psi_\loss$, represents the penalty for not observing $\X_\mis$. It implies that the efficiency loss, $\sigma^2-\sigma^2_\full$, is $E(\psi_\loss^2)$.

\subsection{Debiased machine learning estimation}\label{ssec:debiasedml}

The efficient influence function in Theorem \ref{thm:asy_normal} satisfies the Neyman orthogonality property \citep{ChernozhukovChetverikovDemirerDufloHansenNeweyRobins2018}, which ensures that small perturbations in the estimation of nuisance functions have only second-order effects on the target parameter. This orthogonality motivates using a cross-fitted, debiased machine-learning estimator for $\theta$.

Let $K$ be a fixed positive integer. We randomly partition the sample into $K$ approximately equal folds with index sets $\mathcal{I}_1,\ldots,\mathcal{I}_K$, of sizes $n_1,\ldots,n_K$, respectively. For each fold $k$, we train the nuisance functions using the observations in the set $\mathcal{I}_{-k}=\{1,\ldots,n\}\setminus \mathcal{I}_k$ with sample size $n_{-k}$, and then use these estimates to evaluate the influence function on the held-out fold $\mathcal{I}_k$. This sample-splitting procedure prevents overfitting and ensures valid inference even with highly flexible machine learning methods. We denote by $\mathbb{P}_{n_k}$ and $\mathbb{P}_{n_{-k}}$ the empirical averages over $\mathcal{I}_k$ and $\mathcal{I}_{-k}$, respectively.

We now describe the nuisance estimators obtained from $\mathcal{I}_{-k}$. Specifically, for each missing pattern $\r\in\mathcal{R}\setminus\{\1\}$, we consider the moment condition $\mathcal{A}_{\r}(\delta_{\r})=1$ introduced in Section \ref{ssec:identification}. Noting that $E_{\r}\{\calA_\r(\delta_{\r})-1\}^2/4=\sup_{\phi\in\calL_2(\fZ_\r)} E_{\r}[(\bar{R}\delta_{\r}-1)\phi(\fZ_\r)-\phi(\fZ_\r)^2]$, we propose the following regularized minimax estimator for $\delta_{\r}$:
\begin{align*}
\wh\delta_{\r}^{(k)}=&\amin_{\delta\in\calH}\sup_{\phi\in\calG}  \mathbb{P}_{n_{-k}}\Big[I(\R\in\{\1,\r\})\left\{\left(1-\bar{R}\delta(\fX)\right)\phi(\fZ_{\r})-\phi^2(\fZ_{\r})\right\} \Big]
-\lambda_{\calG}^\delta\Vert \phi\Vert_{\calG}^2+\lambda_{\calH}^\delta\Vert \delta\Vert_{\calH}^2,
\end{align*}
where $\mathcal{H}$ and $\mathcal{G}$ are normed function spaces with norms $\Vert\cdot\Vert_\calH$ and $\Vert\cdot\Vert_\calG$, and $\lambda_{\mathcal{G}}^\delta,\lambda_{\mathcal{H}}^\delta>0$ are regularization parameters. This formulation can accommodate various function classes such as neural networks and reproducing kernel Hilbert spaces (RKHS). Convergence analysis of similar minimax estimators can be found in \citet{DikkalaLewisMackeySyrgkanis2020} and \citet{GhassamiYingShpitserTchetgen2022}. Given $\wh\delta_{\r}^{(k)}$, we construct the estimator of $\omega_0(\fX)$ as
$\wh\omega^{(k)}=1+\textstyle\sum_{\r\in\calR\backslash\{\1\}} \{\wh\delta_{\r}^{(k)}-1\}$. Based on $\wh\omega^{(k)}$, we then estimate $\gamma_0(m,\x)$ and $\eta_0(\x)$ using classical machine learning methods for regression via their identification formulas \eqref{eq:iden_gamma} and \eqref{eq:iden_eta}, yielding estimators $\wh\gamma^{(k)}$ and $\wh\eta^{(k)}$. Next, we obtain estimators for $b_1(\x)$ and $b_2(m,\x)$ sequentially by exploiting the following ratios of conditional mean relationships (see Supplementary Appendix \ref{ssec:app_pf_prop_iden_kappa}):
\begin{align*}
    b_1(\x)=\frac{E\{\omega_0(\fX)\mid \R=\1,\x\}}{E\{(1-T)\omega_0(\fX)\mid \R=\1,\x\}},~~
    b_2(m,\x)=\frac{E\{(1-T)\omega_0(\fX)b_1(\X)\mid \R=\1,m,\x\}}{E\{T\omega_0(\fX)\mid \R=\1,m,\x\}}.
\end{align*}
The corresponding estimates are denoted by $\wh b_1^{(k)}$ and $\wh b_2^{(k)}$. Substituting $(\wh\gamma^{(k)},\wh\eta^{(k)},\wh b_1^{(k)},\wh b_2^{(k)})$ into \eqref{eq:def_r_IF} yields the estimator $\wh\kappa^{(k)}$ of $\kappa$. We next estimate $\nu_\r\triangleq \calA_{\r}(\varrho_{\r})$. Analogous to the construction of $\wh\delta_{\r}^{(k)}$, we consider the following estimator, motivated by the representer equation in Assumption \ref{ass:representer}:
\begin{small}
\begin{align*}
\wh\nu_\r^{(k)}=&\amin_{\nu\in\calG'}\sup_{\delta\in\calH'}\mathbb{P}_{n_{-k}}\Big[\bar{R}\{\wh\kappa^{(k)}-\nu(\fZ_\r)\}\delta(\fX)- \bar{R}\delta^2(\fX) \Big] -\lambda_{\calH'}^\nu\Vert \delta\Vert_{\calH'}^2+\lambda_{\calG'}^\nu\Vert \nu\Vert_{\calG'}^2,
\end{align*}
\end{small}
where $\mathcal{H'}$ and $\mathcal{G'}$ are normed function spaces with norms $\Vert\cdot\Vert_{\calH'}$ and $\Vert\cdot\Vert_{\calG'}$, and $\lambda_{\calH'}^\nu,\lambda_{\calG'}^\nu>0$ are regularized parameters. 
Finally, the cross-fitted debiased machine learning estimator for $\theta$ is given by
\begin{align*}
\wh\theta_{\cf}=\frac{1}{K}\sum_{k=1}^K \mathbb{P}_{n_k}\bigg[ \bar R\wh\omega^{(k)}(\fX)\wh \kappa^{(k)}(T,M,Y,\X)  +  \sum_{\r\in\calR\backslash\{\1\}} \wh\nu_\r^{(k)} \left\{I(\R\in\{\1,\r\})-\bar{R}\wh\delta_{\r}^{(k)}(\fX)\right\}\bigg].
\end{align*}
The following theorem establishes the asymptotic properties of $\wh\theta_{\cf}$ under regularity conditions similar to those required for debiased machine learning estimators of average treatment effects \citep{ChernozhukovChetverikovDemirerDufloHansenNeweyRobins2018}.
\begin{theorem}\label{thm:dml}
     Suppose that for each $\r\in\mathcal{R}\setminus\{\1\}$ and each  $k$, the following conditions hold:
	\begin{itemize}
		\item [(i)]  $\{\wh\delta_{\r}^{(k)},\wh\gamma^{(k)},\wh\eta^{(k)},\wh b_1^{(k)}, \wh b_2^{(k)},\wh\nu_\r^{(k)}\}$ converges in probability to $\{\delta_{\r},\gamma_0,\eta_0,b_1,b_2,\nu_\r\}$;
        \item[(ii)] $\{|b_1|,|b_2|,|\gamma|,|\eta|,|\wh\gamma^{(k)}|,|\wh\eta^{(k)}|\}<C$ for some constant $C>0$;
        \item[(iii)] $n^{1/2}$-convergence of second-order terms, i.e.,
        \begin{small}
        \begin{align*}
            &\Vert \wh b_1^{(k)}-b_1\Vert_2\left\{
            \Vert \wh \gamma^{(k)}-\gamma\Vert_2 +\Vert \wh \eta^{(k)}-\eta\Vert_2     \right\} + \Vert \wh b_2^{(k)}-b_2\Vert_2 \Vert \wh \gamma^{(k)}-\gamma\Vert_2+    
            \Vert \wh\delta_{\r}^{(k)}-\delta_{\r}\Vert_2\left\{\Vert \wh b_1^{(k)}-b_1\Vert_2\right. \\
            &\left.~~~+ \Vert \wh b_2^{(k)}-b_2\Vert_2 + \Vert \wh \gamma^{(k)}-\gamma\Vert_2 +\Vert \wh \eta^{(k)}-\eta\Vert_2+\Vert \wh\nu_\r^{(k)}-\nu_\r\Vert_2\right\}=o_p(n^{-1/2}).
        \end{align*}
\end{small}
        \end{itemize}
    Then, $\wh\theta_{\cf}$ is asymptotically normal, has the influence function $\psi(\O)$, and achieves the semiparametric efficiency bound.
\end{theorem}

\section{Numerical studies}\label{sec:simulation}

In this section, we conduct simulation studies to demonstrate the finite-sample performance of the proposed SIO estimator under various data generation settings. Additional simulation studies are provided in Supplementary \Cref{sec:app_add_simu}, including the proposed debiased machine learning approach with high-dimensional covariates, missing-at-random scenarios, and comparisons with a trimming approach, to further illustrate the performance of the proposed method.

Suppose $(\varepsilon_1,\varepsilon_1',\varepsilon_2,\varepsilon_2',\varepsilon_3,\varepsilon_4)$ are independent standard normal variables, and let $\Phi$ denote the standard normal distribution function. Let $\X=(X_1,X_2,X_3)$, where $X_j\sim \Phi(\varepsilon_j)$ for $j=1,2$ and $X_3\sim\mbox{Bernoulli}(0.5)$. Assume that $\X_\mis=(X_1,X_2)$ and $X_3$ is fully observed. 
Define $\mbox{expit}(u)=\exp(u)/\{1+\exp(u)\}$. 
The treatment is generated by $T\sim\mbox{Bernoulli}\left\{\mbox{expit}\left( 0.1+X_1^{2}-X_2+0.2X_3\right)\right\}$.
The mediator follows $M= -1 + T - TX_1^2 + 4X_1^2 - \sin(X_2) +X_3 + \varepsilon_3$,
and the outcome follows $Y=-1+T+3TM-1.5M+TX_1+5\sin(X_1) - 2X_2^2+X_3 + \varepsilon_4$. 
We consider two kinds of missing mechanisms. 
In DGP1, the missingness indicators $(R_1,R_2)$ of $\X_\mis$ are drawn from a multinomial distribution with probabilities $(p_{11},p_{10},p_{01},p_{00})$, where $p_{11}=\{1+3\exp(-U)\}^{-1}$, $p_{kl}=\{\exp(U)+3\}^{-1}$, and $U=3X_1+X_2-X_3+0.5T\varepsilon_3+0.1\varepsilon_4$ for $kl\in\{10,01,00\}$. In this setting, $\fX=(T,M,Y,\X)$, and we construct two auxiliary shadow variables $Z_j\sim \Phi(\alpha\varepsilon_j+\sqrt{1-\alpha^2}\varepsilon_j')$ with $\alpha=0.6$ for $j=1,2$. 
In DGP2, $(R_1, R_2)$ follow the same multinomial structure but with $U=1.2+X_1 + X_2^2 - X_3 - 0.5T + TX_1^2$. In this setting, $\fX=(T,\X)$, and we use $(M,Y)$ as shadow variables.
In both DGPs, the average percentages of these missingness patterns are about 58\%, 14\%, 14\%, and 14\%, respectively.

The estimands include the mediation functional (MF) $\theta$, NIE $\xi^{(1)}$,  NDE $\zeta^{(0)}$ and ATE $\tau$.
We use polynomial functions to construct the sieve basis and select the sieve dimensions via the data-driven procedures outlined in \Cref{ssec:sel_tuning}. Details of the construction and a histogram of the selected sieve dimensions are provided in Supplementary \Cref{ssec:app_sieve_dim}.
We refer to our proposed estimator as \texttt{SIO}. 
For comparison, we include three additional methods: (i) \texttt{Oracle}, an estimator that assumes complete knowledge of the underlying missing covariates; (ii) \texttt{MI}, a multiple imputation estimator that imputes missing values using observed variables; and (iii) \texttt{CCA}, the complete-case analysis that confines the analysis to the set of units with no missing values. 
In pursuit of fairness, we employ nonparametric estimation for the three comparative methods following \cite{ImaiKeeleTingley2010}, where nuisances are estimated through classical series estimation \citep{Newey1997}.
Across all these approaches, we replicate 500 simulations at sample sizes of 1000 and 2000.

\begin{table}[]
\renewcommand{\arraystretch}{1.1} 
\caption{Summary of the bias and standard deviation (in parentheses) of different estimators under sample sizes 1000 and 2000. 
}
\label{table:bias}
\small
\centering
\tabcolsep=0pt
\begin{tabular*}{\textwidth}{@{\extracolsep{\fill}}crrrrrrrrr@{\extracolsep{\fill}}}
\toprule%
& \multicolumn{4}{@{}c@{}}{$n=1000$} & \multicolumn{4}{@{}c@{}}{$n=2000$} \\
\cline{2-5}\cline{6-9}%
Methods & \multicolumn{1}{c}{MF} & \multicolumn{1}{c}{NIE}  & \multicolumn{1}{c}{NDE} & \multicolumn{1}{c}{ATE} & \multicolumn{1}{c}{MF} & \multicolumn{1}{c}{NIE}  & \multicolumn{1}{c}{NDE} & \multicolumn{1}{c}{ATE} \\
\midrule
\multicolumn{9}{c}{(a) DGP1: Use auxiliary shadow variables}\\
Oracle & 0.00(0.14) & 0.00(0.11) & 0.00(0.21) & 0.01(0.16) & 0.00(0.10) & 0.00(0.08) & 0.00(0.15) & 0.00(0.12)\\
SIO & -0.01(0.16) & 0.03(0.16) & -0.02(0.23) & 0.01(0.19) & -0.01(0.11) & 0.01(0.11) & -0.02(0.16) & -0.01(0.14)\\
MI & 0.23(0.14) & -0.16(0.11) & 0.26(0.20) & 0.10(0.17) & 0.23(0.10) & -0.16(0.08) & 0.25(0.14) & 0.10(0.12)\\
CCA & 0.71(0.20) & 0.11(0.15) & 0.80(0.29) & 0.91(0.22) & 0.71(0.14) & 0.12(0.10) & 0.79(0.20) & 0.91(0.16)\\
\multicolumn{9}{c}{(b) DGP2: Use $(M,Y)$ as shadow variables}\\
Oracle & -0.01(0.15) & 0.01(0.11) & -0.01(0.21) & -0.01(0.16) & -0.01(0.10) & 0.01(0.07) & -0.00(0.15) & 0.00(0.12)\\
SIO & -0.01(0.16) & -0.01(0.15) & 0.01(0.24) & 0.01(0.18) & 0.00(0.12) & -0.01(0.10) & 0.02(0.17) & 0.01(0.13)\\
MI & -0.18(0.17) & 0.09(0.15) & -0.18(0.23) & -0.08(0.17) & -0.18(0.11) & 0.09(0.10) & -0.17(0.16) & -0.08(0.13)\\
CCA & 0.18(0.20) & -0.07(0.15) & 0.25(0.28) & 0.18(0.22) & 0.18(0.14) & -0.07(0.10) & 0.25(0.20) & 0.18(0.16)\\
\bottomrule
\end{tabular*}
\end{table}

\begin{figure}[h!]
    \centering
    \includegraphics[width=\linewidth]{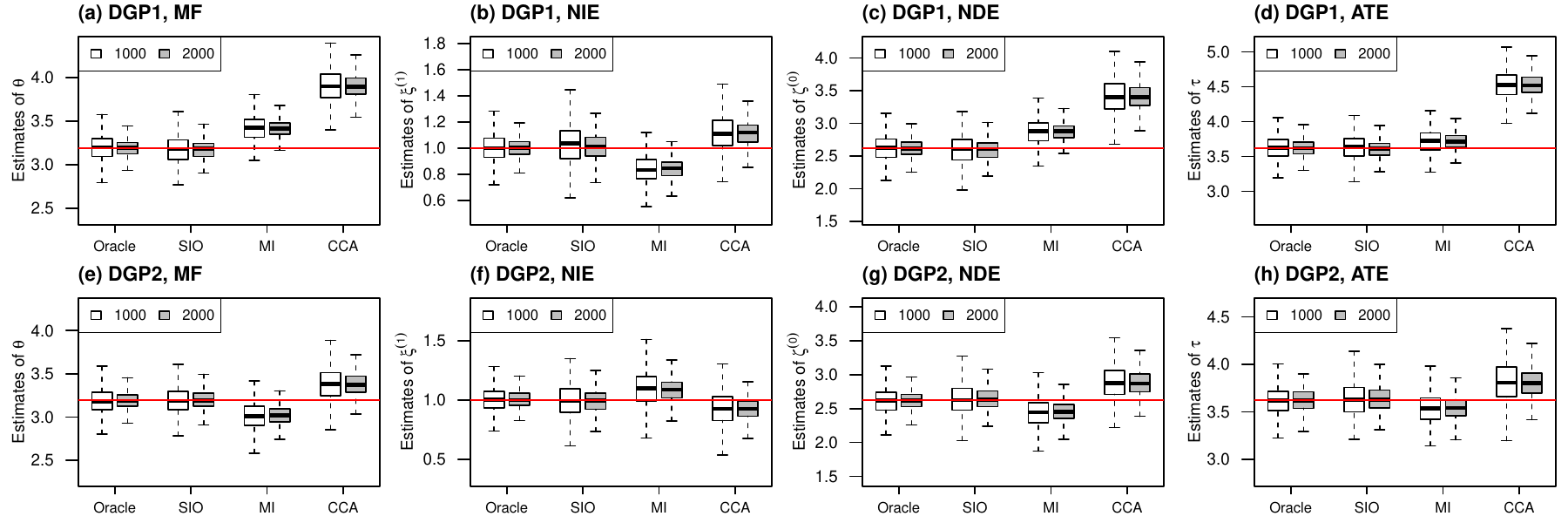}
    \caption{The boxplots of estimates of the different methods among 500 replications under sample sizes 1000 and 2000. The solid line represents the true values.}
    \label{plt:simu1}
\end{figure}

\Cref{table:bias} presents the bias and standard deviation of different estimators, and \Cref{plt:simu1} shows the boxplots over 500 replications. The proposed estimator \texttt{SIO} has negligible bias for all estimands, comparable to that of the \texttt{Oracle} estimator. 
In contrast, both \texttt{MI} and \texttt{CCA} estimators are biased.
While the multiple imputation approach reduces bias relative to complete-case analysis, it is not particularly objective in our setting, as it is typically feasible only for missing-at-random cases and cannot address nonignorable missingness.  The bad performance of \texttt{CCA} further emphasizes the pitfalls of simply discarding observations with missing values in practical applications.
We calculate the 95\% coverage probability (CP) and average length of confidence intervals for various estimators and present results in \Cref{table:cp}.  It is shown that the 95\% CPs of \texttt{SIO} are close to the nominal level. In contrast, both \texttt{MI} and \texttt{CCA} exhibit unsatisfactory CPs due to non-negligible bias. All simulation findings indicate good finite-sample performance for our proposed approach.

\begin{table}[]
  \small
  \centering
  \caption{Summary of the 95\% coverage probability and average length of confidence intervals (in parentheses) of different estimators under sample sizes 1000 and 2000.
  }
  \renewcommand{\arraystretch}{1.1} 
  \label{table:cp}
  \tabcolsep=0pt
  \begin{tabular*}{.9\textwidth}{@{\extracolsep{\fill}}ccccccccc@{\extracolsep{\fill}}}
  \toprule%
  & \multicolumn{4}{@{}c@{}}{$n=1000$} & \multicolumn{4}{@{}c@{}}{$n=2000$} \\
  \cline{2-5}\cline{6-9}%
  Methods & MF & NIE  & NDE & ATE & MF & NIE  & NDE & ATE \\
  \midrule
\multicolumn{9}{c}{(a) DGP1: Use auxiliary shadow variables}\\
Oracle & .968(0.56) & .936(0.40) & .952(0.80) & .960(0.65) & .936(0.40) & .934(0.28) & .950(0.57) & .936(0.46)\\
SIO & .970(0.68) & .956(0.63) & .976(0.96) & .980(0.84) & .968(0.48) & .952(0.45) & .964(0.68) & .968(0.61)\\
MI & .626(0.57) & .604(0.38) & .740(0.79) & .900(0.64) & .400(0.40) & .386(0.27) & .570(0.56) & .840(0.45)\\
CCA & .056(0.74) & .856(0.53) & .182(1.07) & .008(0.87) & .000(0.53) & .774(0.37) & .012(0.76) & .000(0.61)\\
\multicolumn{9}{c}{(b) DGP2: Use $(M,Y)$ as shadow variables}\\
Oracle & .944(0.56) & .928(0.40) & .932(0.80) & .954(0.66) & .938(0.40) & .938(0.28) & .942(0.57) & .954(0.46)\\
SIO & .956(0.65) & .934(0.57) & .946(0.96) & .974(0.78) & .948(0.46) & .956(0.40) & .952(0.68) & .960(0.55)\\
MI & .786(0.62) & .894(0.56) & .854(0.87) & .918(0.68) & .662(0.44) & .862(0.39) & .804(0.62) & .896(0.48)\\
CCA & .824(0.75) & .884(0.53) & .842(1.06) & .868(0.88) & .722(0.53) & .858(0.37) & .734(0.75) & .802(0.62)\\
  \bottomrule
  \end{tabular*}
\end{table}

\section{Empirical application}\label{sec:application}

\label{rev:bi-dir} 
We apply the proposed approach to assess the mediation effects of subjective well-being in the relationship between job satisfaction and depressive symptoms among Chinese adults aged 35--60. The data are from the 2020-2022 wave of the China Family Panel Studies (CFPS), which comprises information on respondents' socioeconomic, demographic, and household-level characteristics.
The outcome $Y$ is the severity of depressive symptoms (from 8 to 32) measured by the Center for Epidemiologic Studies Depression Scale. The treatment is an individual's level of job satisfaction (from 1 to 5), considering six aspects: work income, work safety, work environment, work hours, work promotion and general feelings about work. We assign $T=1$ if the satisfaction level exceeds 3; otherwise, $T=0$. 
The mediator variable $M$ is respondents' subjective well-being (from 1 to 10), measured using a 10-point Likert-type scale question: ``What is your happiness level?''  The covariates $\X$ include age, gender, family size, number of houses, self-reported health, marital status, and income. We select 6519 respondents aged 35-60 who participated in both the 2020 and 2022 surveys as the study sample. To mitigate potential issues of bidirectional causality, we use well-being and depressive symptom measures from the 2022 wave, while treatment and covariates data are drawn from the 2020 wave. This temporal ordering helps establish a more credible causal pathway.\label{data:cfps}

Among the covariates, income is a crucial confounder, as it plays a central role in assessing job satisfaction and is closely associated with well-being and depression. However, about 44\% of the income data are missing. The missingness is likely nonignorable because individuals with high incomes are often less inclined to disclose their earnings.
To apply the proposed method, we consider two choices of shadow variables $\Z$. 
The first is $\Z=(M,Y)$ as in \Cref{exa:mo_independent}.  
The second is an auxiliary variable: the respondent's household Engel coefficient, defined as the percentage of family expenditures spent on food.\label{rev:two-shadow}
This variable is accessible in the CFPS dataset and is well-suited for this role for two reasons. First, Engel's law describes the close relationship between income and food expenditures: households with lower income tend to allocate a higher share of their spending to food than higher-income households. {In our data, the Pearson correlation between the Engel coefficient and income among the observed units is $-$0.11, with a corresponding p-value of $1.5\times10^{-15}$.}  Second, the Engel coefficient is an objective indicator of household income, and it is expected not to directly influence an individual's propensity to conceal income information during the survey among individuals with the same income level and characteristics.

\label{rev:empirical-res} We denote the proposed method using the two types of shadow variables as \texttt{SIO1} and \texttt{SIO2}, respectively. We also use the \texttt{MI} and \texttt{CCA} approaches for comparison. \Cref{table:application} presents the point estimates and 95\% confidence intervals for the NIE, NDE, and ATE. Notably, \texttt{SIO1} and \texttt{SIO2} yield similar results, demonstrating the robustness of the proposed estimator with respect to the selection of shadow variables.
In contrast, the estimates from \texttt{MI} and \texttt{CCA} deviate from those of \texttt{SIO}. In particular, \texttt{CCA} may exclude units with higher incomes whose depressive symptoms are less affected by job status, leading to an overestimated effect on depression.  All methods, however, consistently indicate a significant negative association between job satisfaction and depressive symptoms, with subjective well-being playing a significant mediating role. Specifically, increasing job satisfaction could decrease depressive symptoms by promoting subjective well-being.

\begin{table}[]
  \small
  \centering
  \caption{Empirical results for CFPS dataset.}
  \renewcommand{\arraystretch}{1.1} 
  \label{table:application}
  \begin{tabular*}{.9\textwidth}{@{\extracolsep{\fill}}ccccccc@{\extracolsep{\fill}}}
  \toprule%
  & \multicolumn{2}{@{}c@{}}{NIE} & \multicolumn{2}{@{}c@{}}{NDE}  & \multicolumn{2}{@{}c@{}}{ATE} \\
  \cline{2-3}\cline{4-5}\cline{6-7}%
  Methods & Estimate & 95\% CI  & Estimate & 95\% CI  & Estimate & 95\% CI  \\
  \midrule
  SIO1 & -0.356 & [-0.476,-0.237] & -0.556 & [-0.893,-0.219] & -0.913 & [-1.268,-0.558]\\
  SIO2 & -0.348 & [-0.463,-0.234] & -0.521 & [-0.833,-0.209] & -0.870 & [-1.190,-0.550]\\
  MI   & -0.337 & [-0.411,-0.266] & -0.466 & [-0.660,-0.230] & -0.803 & [-1.008,-0.550]\\
  CCA  & -0.384 & [-0.490,-0.394] & -0.605 & [-0.875,-0.330] & -0.989 & [-1.280,-0.688]\\
  \bottomrule
  \end{tabular*}
\end{table}

\section{Discussion} \label{sec:discussion}
This paper introduces a comprehensive approach to address nonignorable missing confounders in causal mediation analysis. Our framework employs shadow variables and a completeness assumption to identify mediation effects and derives explicit expressions for the target parameters in terms of observed data. We propose a nonparametric estimation framework to mitigate model misspecification and establish large-sample properties while addressing challenges arising from solving a Fredholm integral equation of the first kind. In particular, we show that the resulting estimator is $\sqrt{n}$-consistent and asymptotically normal, and we provide a feasible procedure for inference on mediation effects. The proposed estimator is further shown to be locally efficient, offering new insights into the impact of missing data on causal mediation analysis. Finally, we develop a cross-fitted debiased machine learning approach that enables valid inference even in high-dimensional settings.

This paper opens several avenues for further extension. One promising direction is 
to generalize our analysis beyond binary treatments. While \citet{HuangHuangLintonZhang2024} considered general treatment settings for fully observed data, extending our methodology to accommodate such general treatment regimes under nonignorable missingness is an important problem.\label{rev:huang-compare2}
Another avenue involves settings in which the shadow variables themselves may also be missing. Although our empirical application uses fully observed shadow variables, this may not always hold in practice. In particular, when both confounders and shadow variables are nonignorably missing, identifying and estimating causal effects become substantially more challenging. Developing methods to address such scenarios would further broaden the applicability of shadow-variable-based strategies in real-world studies with complex missing data structures.\label{rev:discussion-res}

\bibliographystyle{agsm}
\bibliography{reference}

\end{document}